\documentclass[11pt,twoside]{article}

\newcommand{\kms}{\mbox{km s$^{-1}$}}
\newcommand{\hgyr}{\mbox{h$^{-1}$Gyr }}

\newcommand{\hkpc}{\mbox{h$^{-1}$kpc }}

\usepackage{asp2004}
\usepackage{epsf}
\usepackage{lscape}
\usepackage{graphicx}

\begin{document}
\title{Simulated Molecular Gas Emission in Galaxy Mergers with
Embedded AGN} %%%

\author{Desika Narayanan$^1$, Thomas J. Cox$^2$, Brant Robertson$^2$,
Romeel Dav\'e$^1$, Tiziana Di Matteo$^3$, Lars Hernquist$^2$, Philip
Hopkins$^2$, Craig Kulesa$^1$, Christopher K. Walker$^1$}

\affil{1: Steward Observatory, University of Arizona, 933 N. Cherry
Ave., Tucson, Az., 85721 2: Harvard-Smithsonian Center for
Astrophysics, 60 Garden St., Cambridge, Ma., 02138, 3: Carnegie Mellon
University, Dept. of Physics, 5000 Forbes Ave., Pittsburgh, Pa.,
15213}
\begin{abstract} 

 We investigate the effect of embedded active galactic nuclei (AGN) in
galaxy mergers on the CO molecular line emission by combining
non-local thermodynamic equilibrium (LTE) radiative transfer
calculations with hydrodynamic simulations. We find that AGN feedback
energy in gas rich galaxy mergers can contribute to large molecular
outflows which may be detectable via velocity-integrated emission
contour maps, as well as through kinematic features in the emission
line profiles.

\end{abstract}

%%% MAIN BODY OF TEXT GOES HERE. CONSULT "INSTRUCTIONS FOR AUTHORS USING
%%% LATEX2E MARKUP", SECTIONS 2.3-2.6 FOR HELP WITH EQUATIONS, FIGURES,
%%% AND TABLES.

%\section{}   %%% Top level section head (remove "%" symbol)
%\subsection{}   %%% Second level section head (remove "%" symbol)
%\subsubsection{}   %%% Lowest level section head (remove "%" symbol)
%\section*{}	%%% Unnumbered top level section head (remove "%" symbol)
%\subsection*{}   %%% Unnumbered second level section head (remove "%" symbol)

\section{Introduction}

Galaxy mergers in the Universe produce a prodigious amount of infrared
luminosity due to dust heating by a combination of the induced
starburst activity, and possibly an enshrouded active galactic nucleus
(AGN). The remnant galaxies, often dubbed ultraluminous infrared
galaxies (ULIRGs, $L_{\rm IR} \geq$10$^{12}$L$_{\sun}$), may represent
a crucial point in an evolutionary sequence of galaxy formation. While
ULIRGs are known to host violent starbursts, they may also serve as
antecedents to quasars. Observations have shown that ULIRGs can
exhibit spectral energy distributions (SEDs) characteristic of both
starburst galaxies, and quasars (Farrah et al. 2003). Recent models
have added a theoretical foundation for the starburst-AGN connection.
Springel, Di Matteo \& Hernquist (2005) and Hopkins et al. (2005a,b)
have shown that there may be a link between the merger induced
starburst and an AGN dominated phase. In this model, feedback energy
associated with the growth of the central black hole(s) (BH) can lift
the veil of obscuring dust and gas, and reveal, along several
sightlines, an optically selected quasar.

There has been a longstanding interest in understanding emission from
the molecular gas in ULIRGs, as the molecular gas serves as the fuel
for the starburst activity, and possibly for accreting BH(s).  Most
ULIRGs in the local Universe have been shown to emit copious molecular
line radiation, both in diffuse and dense form (e.g. Sanders, Scoville
\& Soifer, 1991, Narayanan et al. 2005). Recent studies by Greve et
al. (2005), Hainline et al. (2005), Carilli et al. (2005) and Tacconi
et al. (2006) have shown that IR bright galaxies at high redshift
($z\sim 2$) contain a significant amount of molecular gas as well. As
evidenced by X-ray and infrared (IR) studies, many of these galaxies
at high-$z$ are known to contain AGN (Alexander et al. 2005, Polletta
et al. 2005). In this contribution, we discuss the possible impact of
embedded AGN on the observed molecular line emission in galaxy
mergers.

\section{Numerical Methods: SPH and non-LTE Radiative Transfer}
We have carried out hydrodynamic simulations with an improved version
of the smooth particle hydrodynamics (SPH) code, GADGET-2 (Springel,
2005; Springel et al 2005). The code includes a
prescription for AGN feedback in which a fraction (0.05 \%) of the
accreted mass energy onto the central BH(s) is reinjected into the ISM
as thermal energy. The code allows for radiative cooling of the
gas, a multi-phase structure of the ISM which includes cold clouds
embedded in a hot pressure confining ISM, and a prescription for star
formation constrained by the Schmidt/Kennicutt laws (Kennicutt, 1998;
Schmidt, 1959; see also Springel \& Hernquist, 2003).

The progenitor galaxies in the merger simulation presented in this
contribution are roughly Milky-way like, but with 50\% gas fraction.
They are thus likely reasonable representatives of the progenitors of
present-day ULIRGs. The dark matter halos were initialized with a 
Hernquist (1990) profile and the galaxies followed a parabolic orbit,
initially separated by 140 kpc. In order to simplify the analysis
presented here, we have not included supernovae winds in our models.

Our non-LTE radiative transfer methods are based on an improved
 version of the Bernes (1979) Monte Carlo algorithm which we more
 fully describe in Narayanan et al. (2006a,b). Our improvements focus
 on including giant molecular clouds (GMCs) as singular isothermal
 spheres in a sub-grid fashion in order more accurately model emission
 from high excitation lines which originate in the dense cores of
 molecular clouds. In this study we consider the CO molecular line
 emission and use $\sim$10$^7$ model photons per iteration.

\section{Results}

\subsection{Morphology}
\label{section:morphology}

\begin{figure}

\includegraphics[angle=90,scale=0.45]{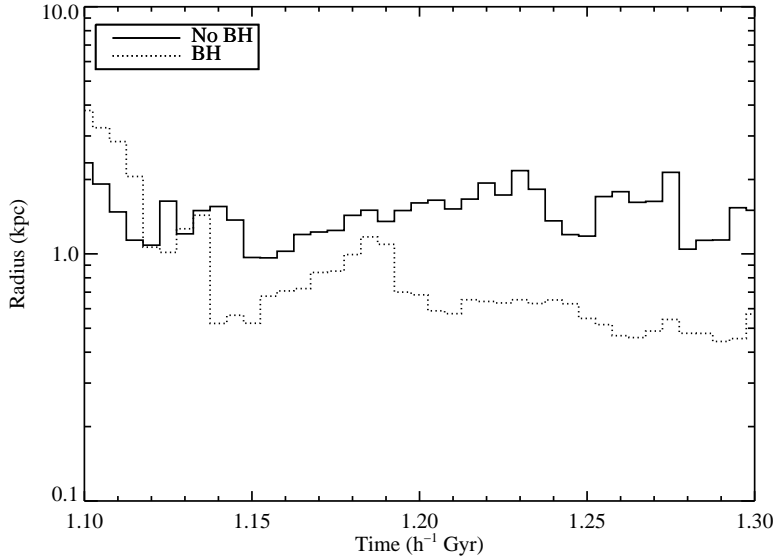}
\caption{CO (J=1-0) half light radius as a function of time for models
of galaxy mergers without (solid line) and with (dash line) black
holes.  The AGN winds in the black hole model remove the extended
diffuse gas around the nucleus, leaving behind only a compact core of
dense molecular gas.\label{figure:halflight}}
\end{figure}

We have run two identical simulations, one with BHs, and one
without. While the CO morphologies in these galaxies generally follow
similar evolutionary paths, there are some distinct differences
arising from thermal energy input by the accreting BH(s).

In the merger simulation without BHs, after the galaxies merge, the
star formation proceeds on scales of a few kiloparsecs.  The starburst
continues, relatively unhindered, until most of the cold molecular gas
is consumed (Springel et al. 2005). In contrast, when BHs are included
in the simulations, the star formation becomes compact soon after the
progenitor galaxies coalesce. An AGN driven wind blows the diffuse gas
out of the nucleus, leaving behind only a dense core of star forming
gas, typically confined to the central 500-1000 pc.  These phenomena
are quantified in Figure~\ref{figure:halflight}, where we show the CO
(J=1-0) half-light radius as a function of time for both
simulations. Observationally, CO emission is seen to be heavily
concentrated within the central $\sim$1/2 kpc in advanced mergers
(e.g. Bryant \& Scoville, 1999).

\begin{figure}
\includegraphics[scale=0.575]{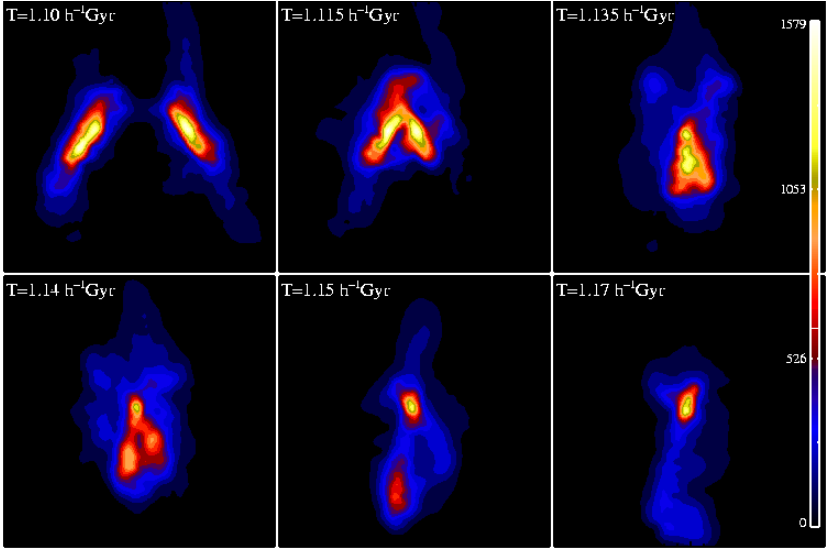}
\caption{CO (J=1-0) emission contours for an equal mass gas rich
galaxy merger. After the galaxies merge, gaseous inflows fuel black
hole growth. Subsequent AGN feedback energy can blow large amounts of
molecular gas out from the nuclear regions. The time stamp of each
image is in the top left of each panel, and the intensity contours are
in units of K-\kms.  The scale for the contours is on the right. Each
panel is 12 \hkpc on a side, and the plots are made at 1/4 kpc spatial
resolution. \label{figure:outflowmaps}}
\end{figure}

During the period of heaviest black hole accretion, the AGN feedback
energy can have a dramatic effect on the CO morphology.  In
Figure~\ref{figure:outflowmaps}, we show CO (J=1-0) emission contours
for a series of snapshots from our hydrodynamic simulation with BHs
spanning $\sim$60 h$^{-1}$ Myr. The plot traverses a time series
beginning when the galaxies are just approaching for final coalescence
through the beginning of the phase of maximum BH accretion/feedback.
As is evident from T$\sim$1.14 h$^{-1}$ Myr onward, the galaxy system
ejects large columns of molecular gas through an AGN-driven wind.
These outflows may be imageable via CO emission. While in detail the
AGN wind may destroy some clouds owing to thermal conduction, enough
gas is entrained in the winds such that a large column of molecular
gas survives. These outflows should be imageable with high-density
molecular gas tracers as well. Our simulations show significant CO
emission from the outflows in the CO (J=3-2) and CO (J=6-5)
transitions. Morphological signatures similar to those see in
Figure~\ref{figure:outflowmaps} may have been observed in LIRG NGC 985
(Appleton et al. 2001). The simulations we ran without BHs do not show
molecular outflows. Additional simulations are needed to investigate
whether supernovae winds can drive similar columns of gas.

\subsection{Line Profiles}

Large scale outflows due to AGN activity may be observable even with
low spatial resolution observations through their imprint on the
emission line profile.  Typically, once the galaxies have merged (in
our simulations, when the BHs are no longer individually
distinguishable), the characteristic CO emission line profile is well
described by a single Gaussian, centered at the systemic velocity of
the galaxy.  However, when viewing outflows with a significant line of
sight (LOS) component, a secondary peak can appear in the line
profile, red or blueshifted at the LOS velocity of the outflow.

\begin{figure}[t]
\includegraphics[angle=90,width=400pt,height=280pt]{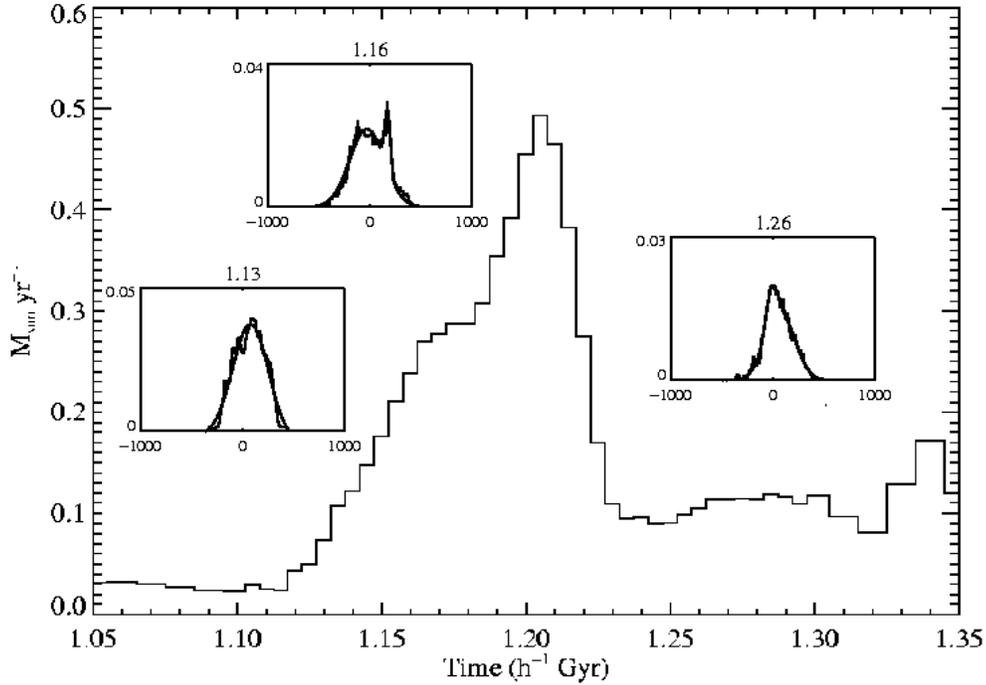}
%\plotone{junk.ps}
\caption{Black hole accretion rate as a function of time for the model
with BHs and selected CO (J=1-0) emission line profiles. The line
profiles are modeled such that the outflow of
Figure~\ref{figure:outflowmaps} is moving away from the observer,
along the line of sight. The time associated with each line is above
the spectra. The high velocity narrow peak in the emission line at
T$\sim$1.16\hgyr is indicative of outflowing gas along the line of
sight. The y-axis for the spectra is Rayleigh-Jeans temperature, and
the x-axis is offset velocity in \kms. The best double-Gaussian fit
for each spectra is over plotted in dashed
lines. \label{figure:lineprofile} }
\end{figure}

In Figure~\ref{figure:lineprofile}, we show the BH accretion rate as a
function of time and a sample of CO (J=1-0) emission line profiles. In
an effort to model the spectral emission from an unresolved
observation, we have set the merger model at $z$=2
($\Omega_\Lambda$=0.7, $\Omega_M$=0.3, $h$=0.75), and convolved our
simulated observations with a 30$\arcsec$ single-dish Gaussian
beam. The emission line profiles are modeled such that the observer is
viewing the outflow of Figure 2 moving away from them with a large LOS
velocity component. Near T=1.16 \hgyr, a high velocity narrow peak
appears in the emission line.  This line profile of a broad Gaussian
with a narrow outflow-induced peak appears to be characteristic of
outflowing gas along the line of sight in our models. We estimate the
``outflow'' profile to be visible $\sim$25\% of the time, averaged
over many viewing angles.

Double peaked profiles are known to exist in mergers in cases that do
not necessarily correspond to outflows. Mergers prior to final
coalescence exhibit a double peak where the emission peaks originate
in the starburst nuclear regions of the progenitor galaxies. Similar
profiles also arise in galaxies in which there is significant
rotation. The degeneracy between the double-peak or rotation profile,
and the ``outflow `` profile may be removed from the shape of the
actual lines, however.  From our simulations, the double peak
corresponding to molecular outflows is always a broad Gaussian with a
narrow peak red or blueshifted from the systemic velocity of the
galaxy. The narrow peak can be quite bright due to the high columns
along the line of sight, but typically much narrower than the $\ga$500
km s$^{-1}$ characteristic of the accompanying broad-line component
owing to small velocity dispersions in the outflow. Conversely, the
line profiles originating in an inclined rotating system or from a
pre-coalescence merger are typically represented by two broad,
symmetric Gaussians, centered about the systemic velocity of the
galaxy, and do not have a narrow line component as seen in our
simulations with outflows.

\begin{figure}
\includegraphics[scale=0.75]{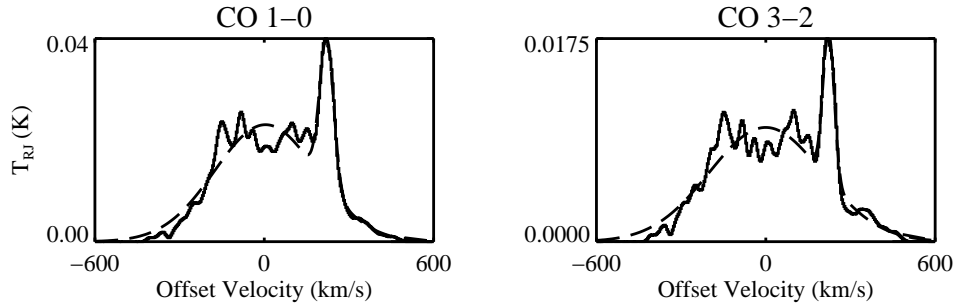}
\caption{Emission line profiles in CO (J=1-0) and (J=3-2), viewing the
T=1.15 \hgyr snapshot of Figure 2 such that the outflow is along the
line of sight. The high velocity peak is due to outflowing gas, and is
visible even in high excitation gas. The best double-Gaussian fit for
each spectra is over plotted in dashed lines.\label{figure:excite}}
\end{figure}

The outflows can exhibit high levels of molecular
excitation. In Figure~\ref{figure:excite}, we present the emission
line profile viewing the outflow in Figure~\ref{figure:outflowmaps}
moving along the line of sight in CO (J=1-0) and CO (J=3-2).  CO
(J=1-0) traces most molecular gas, whereas CO (J=3-2) preferentially
is excited in the dense cores of molecular clouds. The component of
the emission line due to the gas entrained in the AGN wind remains
bright through many high-$J$ transitions, suggesting a possible
relevance to observed emission line profiles in high-$z$ mergers.

\acknowledgements %%% Text of acknowledgements runs on after this command.

  D.N. was funded for this work by an NSF Graduate Research Fellowship.

%%% THE BIBLIOGRAPHY
%%%
%%% CONSULT SECTION 3 OF "INSTRUCTIONS FOR AUTHORS" FOR HOW TO USE NATBIB.
%%% AUTHORS ARE ENCOURAGED TO USE EITHER THE "THEBIBLIOGRAPY" ENVIRONMENT
%%% BY UNCOMMENTING (DELETING THE "%" SYMBOL) THE COMMANDS BELOW, OR BY
%%% USING THE BIBTEX ENVIRONMENT. TO FIND OUT WHICH IS APPLICABLE TO YOUR
%%% CONTRIBUTION, CONSULT THE VOLUME EDITORS FOR YOUR PROCEEDINGS.
%%%

%\includegraphics[angle=90,scale=0.75]{narayanan_d_fig2.eps}

\end{document}